\documentstyle[12pt,aasms4,epsf]{article}	

\def\nTBmax   {\wisk{1.17^{+0.34}_{-0.41}}}
\def\nTBng    {\wisk{1.22^{+0.60}_{-0.71}}}
\def\nTBcross {\wisk{1.02^{+0.44}_{-0.47}}}
\def\nTBDth   {\wisk{1.13^{+0.3}_{-0.4}}}
\def\Qmax     {\wisk{17.9^{+1.6}_{-1.6}}}
\def\Qng      {\wisk{19.6^{+2.9}_{-2.6}}}
\def\Qcross   {\wisk{17.3^{+1.6}_{-2.1}}}
\newcommand{\MS}{\mbox{$\!\!\!$}}

\newif\ifms
\mstrue

\def\wisk#1{\ifmmode{#1}\else{$#1$}\fi}

\def\Amp     {\wisk{{\langle Q_{RMS}^2\rangle^{0.5}}}}

\def\etal    {{\rm et al.}}

\newcommand{\bc}{\begin{center}}
\newcommand{\ec}{\end{center}}

\newcommand{\be}{\begin{equation}}
\newcommand{\ee}{\end{equation}}

\slugcomment{Submitted to the {\it Astrophysical Journal}}

\def\COBE{{\em COBE}}

\begin{document}

\title{Angular Power Spectrum of the Microwave Background Anisotropy
seen by the\\ 
\COBE\footnotemark[1] Differential Microwave Radiometer}

\footnotetext[1]{The National Aeronautics and Space Administration/Goddard
Space Flight Center (NASA/GSFC) is responsible for the design, development,
and operation of the Cosmic Background Explorer (\COBE).
Scientific guidance is provided by the \COBE\ Science Working Group.
GSFC is also responsible for the development of the analysis software
and for the production of the mission data sets.}

\author{
E.~L.~Wright\altaffilmark{2},
C.~L.~Bennett\altaffilmark{3},
K.~G\'orski\altaffilmark{4},
G.~Hinshaw\altaffilmark{4},
\&
G.~F.~Smoot\altaffilmark{5}}

\altaffiltext{2}{UCLA Astronomy, P.O. Box 951562, Los Angeles CA 90095-1562
(I: wright@astro.ucla.edu)}
\altaffiltext{3}{NASA Goddard Space Flight Center, Code 685, Greenbelt MD 20771}
\altaffiltext{4}{Hughes/STX Corporation, 
Laboratory for Astronomy and Solar Physics, 
Code 685, NASA/Goddard Space Flight Center, Greenbelt, Maryland 20771}
\altaffiltext{5}{Dept. of Physics, CfPA, LBL \& SSL, Bldg 50-351,
Univ. of California, Berkeley CA 94720}

\begin{abstract}
The angular power spectrum estimator developed by
Peebles \markcite{P73} (1973) and
Hauser \& Peebles \markcite{HP73} (1973)
has been modified and applied to the 4 year maps
produced by the \COBE\ DMR.  The power
spectrum of the observed sky has been compared to the power
spectra of a large number of simulated random skies
produced with noise equal to the observed noise
and primordial density fluctuation power spectra
of power law form, with $P(k) \propto k^n$.
The best fitting value of the spectral index
in the range of spatial scales
corresponding to spherical harmonic indices $3 \leq \ell \lesssim 30$
is an apparent spectral index $n_{app} = \nTBDth$
which is consistent with the Harrison-Zel'dovich primordial
spectral index $n_{pri} = 1$
The best fitting amplitude for $n_{app} = 1$ is $\Amp = 18\;\mu$K.
\end{abstract}

\section{Introduction}

The spatial power spectrum of primordial density perturbations,
$P(k)$ where $k$ is the spatial wavenumber, provides evidence
about processes occurring very early in the history of the Universe.
In the first moments after the Big Bang, the horizon scale $ct$
corresponds to a current scale that is much smaller than galaxies,
so the assumption of a scale free form for $P(k)$ on large scales
is natural,  which implies a power law $P(k) \propto k^n$.
The Poisson equation $\nabla^2\phi = 4\pi G \rho$ implies
that $\vert\phi(k)\vert^2 \propto P(k)/k^4$, and hence that the potential
fluctuations on scale $\lambda \propto 1/k$ are
$\Delta \phi^2 \propto \vert\phi(k)\vert^2 k^3 \propto k^{n-1}$.
Harrison \markcite{H70} (1970), 
Zel'dovich \markcite{Z72} (1972), 
and Peebles \& Yu \markcite{PY70} (1970) all
pointed out that the absence of tiny black holes implies 
that the limit as $k \rightarrow \infty$ of $\Delta \phi$ must be small,
so $n \lesssim 1$,
while the large-scale homogeneity implied by the near isotropy of the
Cosmic Microwave Background Radiation (CMBR) requires 
that the limit as $k \rightarrow 0$ of $\Delta \phi$ must be small,
so $n \gtrsim 1$.
Thus the prediction of a Harrison-Zel'dovich or $n = 1$ form for $P(k)$
by an analysis that excludes all other possibilities is an old one.
This particular scale-free power law is scale-invariant because the
perturbations in the metric (or gravitational potential) are independent
of the scale.
The inflationary scenario 
(Starobinsky \markcite{S80} 1980; Guth \markcite{G81} 1981) 
proposes a tremendous expansion
of the Universe (by a factor $\gtrsim 10^{30}$)
during the inflationary epoch, which can convert
quantum mechanical fluctuations on a microscopic scale during the 
inflationary epoch into Gpc-scale structure now.  To the extent that
conditions were relatively stable during the small part of the
inflationary epoch which produced the Mpc to Gpc structures we now study,
an almost scale-invariant spectrum is produced 
(Bardeen, Steinhardt \& Turner \markcite{BST83} 1983).
Bond \& Efstathiou \markcite{BE87} (1987)
show that the expected variance of the coefficients
$a_{\ell m}$ in a spherical harmonic expansion of the CMBR temperature
given a power law power spectrum $P(k) \propto k^n$ is 
$<a_{\ell m}^2> \; \propto \Gamma[\ell+(n-1)/2] / \Gamma[\ell+(5-n)/2]$
for $\ell < 40$.
Thus a study of the angular power spectrum of the CMBR can be used to place
limits on the spectral index $n$ and test the inflationary prediction
of a spectrum close to the Harrison-Zel'dovich spectrum with $n = 1$.

The angular power spectrum contains the same information as the 
angular correlation function, 
but in a form that simplifies the visualization of
fits for the spectral index $n$.  
Furthermore, the off-diagonal elements of the covariance matrix
have a smaller effect for the power spectrum than for the correlation
function.
However, with partial sky coverage the multipole estimates in the
power spectrum are correlated, and this covariance must be considered
when analyzing either the correlation function or the power spectrum.

The power spectrum of a function mapped over the entire sphere can be
derived easily from its expansion into spherical harmonics, but for a
function known only over part of the sphere this procedure fails.
Wright \etal \markcite{WSBL94} (1994) have modified a power spectral 
estimator from Peebles \markcite{P73} (1973)
and Hauser \& Peebles \markcite{HP73} (1973) 
that allows for partial coverage
and applied this estimator to the DMR maps
of CMBR anisotropy.
We report here on the application of these statistics to the
DMR maps based on all 4 years of data 
(Bennett \etal \markcite{FourYear} 1996).
Monte Carlo runs have been used to 
calculate the mean and covariance of the power spectrum.
Fits to estimate \Amp\ and $n$ by maximizing the Gaussian approximation
to the likelihood of the angular power spectrum are discussed in this paper.
Since we only consider power law power spectrum fits in this paper,
we use $Q$ as a shorthand for \Amp\ or $Q_{rms-ps}$, 
which is the RMS quadrupole averaged over the whole Universe,
based on a power law fit to many multipoles.
\Amp\ should not be confused with the actual quadrupole of the 
high galactic latitude part of the sky observed from the
Sun's location within the Universe,
which is the $Q_{RMS}$ discussed by 
Bennett \etal\markcite{BSHWK92} (1992).

\section{Estimating the Angular Power Spectrum}

Wright \etal\markcite{WSBL94} (1994)
have discussed the modification of the Hauser-Peebles
angular power spectrum estimator for use on CMBR anisotropy maps.
To allow for the cutting out of the galactic plane, new basis functions
are defined using a modified inner product:
\be
<fg> = {{\sum_{j=1}^N w_j f_j g_j }\over {\sum_{j=1}^N w_j }}
\ee
where $j$ is an index over pixels, and $w_j$ is the weight per pixel.
In the galactic plane cut, $w_j = 0$.  We have {\em not} used weights
proportional to the number of observations, so $w_j = 1$ outside of
the galactic plane cut.  The custom galactic cut used in this paper
basically follows $\sin |b| = 1/3$ with extra cut added in Sco-Oph and
Orion (Banday \etal\markcite{Designer} 1996).
A total of 3881 pixels are used, or 63\% of the sky.

The modified Hauser-Peebles method in 
Wright \etal\markcite{WSBL94} (1994)
used basis functions defined using
\be
G_{\ell m} = F_{\ell m}
 - {{F_{00}<F_{00}F_{\ell m}>}\over{<F_{00}F_{00}>}}
 - \sum_{m^\prime=-1}^1 {{F_{1m^\prime}<F_{1m^\prime}F_{\ell m}>}
                              \over{<F_{1m^\prime}F_{1m^\prime}>}}
\label{MD}
\ee
where the $F_{\ell m}$ are real spherical harmonics and the inner product
$<fg>$ is defined over the cut sphere.  These functions $G_{\ell m}$
are orthogonal to monopole and dipole terms on the cut sphere.
Call this the MD
method since the basis functions are orthogonal to the monopole and dipole.
Let the MDQ method use basis functions orthogonal to the monopole, dipole
and quadrupole:
\be
G^\prime_{\ell m} = F_{\ell m}
 - {{F_{00}<F_{00}F_{\ell m}>}\over{<F_{00}F_{00}>}}
 - \sum_{m^\prime=-1}^1 {{F_{1m^\prime}<F_{1m^\prime}F_{\ell m}>}
                              \over{<F_{1m^\prime}F_{1m^\prime}>}}
 - \sum_{m^\prime=-2}^2 {{F_{2m^\prime}<F_{2m^\prime}F_{\ell m}>}
                              \over{<F_{2m^\prime}F_{2m^\prime}>}}.
\label{MDQ}
\ee

In this paper we have used the MDQ method so our results
for $\ell \geq 3$ are completely independent of the quadrupole in the map.
We have also tabulated the power in $\ell = 2$ which is computed with
$G^\prime_{2m}$'s which are orthogonal to the monopole, dipole, 
and those components of the quadrupole which occur earlier in the
sequence than $m$.
Because the galactic cut used is not a straight $|b|$ cut, the
different $F_{2m}$'s are not quite orthogonal, and the definition of
$G^\prime_{2m}$ depends slightly on the ordering of the $F_{\ell m}$'s.
We use the ordering $1,\;\cos\phi\,\;\cos 2\phi,\;\sin\phi,\;\sin 2\phi$.
With these basis functions we compute the power spectrum estimators
\be
{{T_\ell^2}\over{2\ell+1}} = 
{ {\sum_{m=-\ell}^\ell <G^\prime_{\ell m}T>^2} \over 
  {\sum_{m=-\ell}^\ell <G^\prime_{\ell m}G^\prime_{\ell m}>} }
\ee
which are quadratic functions of the maps.
Note that for full sky coverage, $T_\ell^2$ is the variance of the sky in
order $\ell$, but for partial sky coverage the response of $T_\ell^2$
to inputs with $\ell^\prime \neq \ell$ causes $T_\ell^2$ to be larger
than the order $\ell$ sky variance.  Table 1 of 
Wright \etal\markcite{WSBL94} (1994) shows the input-output matrix for
a straight $20^\circ$ cut, while Table \ref{tab.io} shows the input-output
matrix for the custom galaxy cut.
The jump in Figure \ref{fig.linear} at $\ell = 5$ for the mean spectrum
of $Q = 17\;\mu\mbox{K},\;n = 1$ inputs is caused by the off-diagonal
response to $\ell = 3$, while the off-diagonal response of $\ell = 4$
to $\ell = 2$ has been zeroed by the MDQ method.

This method computes the power spectrum, a quadratic function
of the map, which includes contributions from both the true
sky signal and from instrument noise.  
We remove the contribution of the instrument noise by subtracting
the power spectrum of a noise only map.  This difference map can be 
constructed
by subtracting the two maps made from the $A$ and $B$ sides of the DMR
instruments: $D = (A-B)/2$.  The sum map containing the real signal is
$S = (A+B)/2$.  When we compute the quadratic power spectrum, we use the
value $S^2 - D^2 = A \times B$, the power spectra reported here are the
cross power spectra between the $A$ and $B$ sides of the DMR instrument.

We have computed the power spectrum of the
internal linear combination ``free-free free'' no galaxy (NG) map
(Wright \etal\markcite{WSBL94} 1994),
$T_{NG} = -0.4512 T_{31} + 1.2737 T_{53} + 0.3125 T_{90}$,
and the close to maximum signal-to-noise ratio maps
based on $0.6 T_{53}/0.931 + 0.4 T_{90}/0.815$.
The denominators in the latter expression convert the Rayleigh-Jeans 
differential temperatures $T_{53}$ and $T_{90}$
into thermodynamic $\Delta T$'s, but this conversion is included in
the coefficients for $T_{NG}$.
This process can also be applied to compute the cross power spectrum of
the 53~GHz and 90~GHz maps by letting $S = (53+90)$ and $D = (53-90)$,
after both maps have been converted into thermodynamic brightness 
temperature differences.  Figure \ref{fig.linear} shows the resulting
power spectra for the three map combinations.

We have binned the power spectra in quasi-logarithmic bins in $\ell$
in order to minimize the increasing noise-to-signal ratio as $\ell$
gets large.  Figure \ref{fig.logpower} 
and Table \ref{tab.logpower} show these binned power spectral estimates.

\section{Monte Carlo Simulations}

In order to calibrate and test these methods for biases, it is necessary to
simulate both the {\it cosmic variance}, which gives a random map 
with random spherical harmonic amplitudes chosen from a Gaussian
distribution with a variance determined from the chosen $Q_{in}$ and
$n_{in}$, and the {\it experimental variance}, which gives the
360 million noise values needed per year.  While programs to simulate
the DMR time-ordered data do exist, 
we have not worked at this level of detail.  
Instead, we have used simulations that start with the maps.

The effect of noise on the map production process can be simulated using
\be
T_n = \sigma_1 A^{-0.5} U
\ee
where $\sigma_1$ is the noise in one observation,
$U$ is an uncorrelated vector of unit variance zero mean Gaussian
random variables, and $A$ is the matrix with diagonal elements
$A_{ii}$ equal to the
number of times the $i^{th}$ pixel was observed, and off-diagonal elements
$-A_{ij}$ equal to the number of times the $i^{th}$ pixel was referenced
to the $j^{th}$ pixel.
Even though $A$ is singular, Wright \etal\markcite{WSBL94} (1994)
give a rapidly convergent
series technique for generating noise maps.  Thus each noise map
depends on 6144 independent Gaussian unit variance random variables
and the parameter $\sigma_1$.

The signal map that is added to the noise maps 
to give the ``observed'' maps is generated using independent
Gaussian random amplitudes with variances given by
Bond \& Efstathiou \markcite{BE87} (1987)
for $\ell < 40$.
The simulations done here included $\ell$'s up to 39, so the signal map 
depends on 1600 Gaussian independent unit variance random variables and 
the two parameters $Q_{in}$ and $n_{in}$.

\section{Maximum Likelihood Estimation}

Using the Monte Carlo's, we find 
the mean power spectrum $\overline{T_\ell^2(Q_{in},n_{in})}$,
and the covariance matrix 
$C(Q_{in},n_{in})_{\ell\ell^\prime} = 
\langle (T_\ell^2 - \overline{T_\ell^2})
(T_{\ell^\prime}^2 - \overline{T_\ell^2}) \rangle$.
For the actual power spectrum $T_\ell^2$ from the real sky or a Monte
Carlo simulation, define
the deviation vector $e_\ell = T_\ell^2 - \overline{T_\ell^2(Q_{in},n_{in})}$
and the $\chi^2$ statistic $\chi^2 = e^T C^{-1} e$.
All of the fits in this paper are based on the range 
$\ell = \ell_{min}\ldots\ell_{max}$ with $\ell_{min} = 3$ and
$\ell_{max} = 30$.  
$C$ is thus a $28 \times 28$ matrix.  
Ignoring the quadrupole is reasonable because the galactic corrections
are largest for $\ell = 2$, and the maximum order used is set by the
DMR beam-size of 7$^\circ$ and the increased computer time required
to analyze more orders.
Since the magnitude of the covariance matrix 
gets larger rapidly when $Q_{in}$ increases 
there is a bias toward large values of $Q$ when minimizing $\chi^2$.
One can allow for this by minimizing $-2 \ln(L)$ instead of $\chi^2$,
where $L$ is the Gaussian approximation to the likelihood:
\begin{equation}
-2 \ln(L) = \chi^2 + \ln(\det(C)) + {\rm const}.
\label{gausslike}
\end{equation}

For any given power spectrum, we can adjust $Q_{in}$ and $n_{in}$ until
$-2 \ln L$ is minimized.  This gives us the maximum likelihood fit of
a power law power spectrum to the given power spectrum.
We have called
the values of $Q$ and $n$ that maximize the likelihood for the
observed power spectrum $Q_{ML}$
and $n_{ML}$ since these are maximum likelihood values.
The maximum likelihood technique gives an {\it asymptotically} unbiased
determination of the amplitude $Q$ and index $n$, but only as the
observed solid angle goes to infinity.  Since we are limited to about
8~sr of sky, asymptotically unbiased means {\it biased} in practice,
both for the quadratic statistics considered here and for linear
statistics used by G\'orski \etal\markcite{GHBBW94} (1994)
and Bond \markcite{B95} (1995).
Our use of a Gaussian approximation to the likelihood
for our quadratic statistics can introduce additional errors.
We use our Monte Carlo simulations to calibrate our statistical
methods to avoid biased final answers.

Our $k^{th}$ Monte Carlo run depends on a set of random variables $\{Z_k\}$
(with 1600 + 12288 elements for a cross-analysis needing two noise maps)
having a known distribution, and the three parameters $Q_{in}$, $n_{in}$
and $\sigma_1$.  $\sigma_1$ can be determined with great precision
using the time-ordered data. 
Hence one needs to run many Monte Carlo simulations with
several different values for $Q_{in}$ and $n_{in}$ and compare
the fitted values $Q_{ML,k}$ and $n_{ML,k}$ to the fitted values for the
real data, $Q_{ML,obs}$ and $n_{ML,obs}$.  
For the $k^{th}$ realization $\{Z_k\}$,
the fitted values $Q_{ML,k}$ and $n_{ML,k}$ are a continuous function
of the input parameters $Q_{in}$ and $n_{in}$, and one can choose
values $Q_{in} = Q_{match,k}$ and $n_{in} = n_{match,k}$ 
such that $Q_{ML,k} = Q_{ML,obs}$
and $n_{ML,k} = n_{ML,obs}$.  By choosing many different realizations of
$\{Z_k\}$, one creates many different $(Q_{match},\;n_{match})$ pairs.
Figure \ref{fig.match} shows this cloud of points for the 2 year
$53 \times 90$ cross-power spectrum, along with contours of
$-2 \ln L$.
The spectral index we give is the median of the set of $n_{match}$'s,
and the 16\%-tile to 84\%-tile range in $n_{match}$ defines the
$\pm 1 \sigma$ range.
The value \nTBDth\ given in the Abstract is the weighted mean of
these determinations, but we have not reduced the error because
the cosmic variance is common to all three maps.

The value of \Amp\ can be found by doing a one parameter maximum
likelihood fit for $Q_{ML}$ with $n$ fixed at 1.  After using
the Monte Carlo runs to debias the maximum likelihood results,
we get the values shown in Table \ref{tab.Qn}.
We can also find the best fit values of $Q$ for other values of $n$.
The best fit $Q$ values for $n$ forced to be 1.25 are smaller than
those for $n$ forced to be 1 by an amount that allows us to estimate
the effective wavenumber of our amplitude determination.  We find
that $\ell_{eff} = 7.9$ for the 53+90 $A\times B$ case and
7.3 for the $53 \times 90$ case.  In Figure  \ref{fig.match} we have 
chosen to plot the amplitude at $\ell = 7$ which is closest to the effective
wavenumber in order to minimize the correlation between the amplitude 
and the spectral index.

\section{Discussion}

The angular power spectrum of the four year \COBE\ DMR maps has
been calculated, and it is very consistent with a
Harrison-Zel'dovich primordial spectrum $n_{pri} = 1$, especially
after the small correction for the ``toe'' of the Doppler peak
which gives an expected apparent index of $n_{app} \approx 1.1$
for $\Omega = 1$ CDM models.
Models with a cosmological constant ($\Lambda$CDM) predict a
smaller $n_{app} \approx 0.75$ 
(Kofman \& Starobinsky \markcite{KS85} 1985) 
that is still consistent with the \COBE\ DMR observations.
The amplitude derived from this analysis is in between the
$\Amp = 17\;\mu$K derived from the first year maps and
the $\Amp = 19\;\mu$K derived from the two year maps
(Wright \etal\markcite{WBSL94} 1994).  The amplitude from the
no galaxy map is consistent with but now slightly higher
than the amplitude derived from the 53 and 90 GHz maps,
indicating that galactic contamination is not a major problem
with the chosen galactic cut.

\acknowledgements
We are grateful for the efforts of the \COBE\ team and the support of
the Office of Space Sciences at NASA.  Charley Lineweaver provided helpful
comments on an early draft of this paper.

\clearpage


\begin{deluxetable}{rrrrrrrrrrrrrrrrrrr}
\scriptsize
\tablewidth{0pt}
\tablecaption{
$10^3 \times$ the mean over $m$ of the power spectra of 
$F_{\ell^\prime m}$ for the custom galaxy cut.\label{tab.io}}
\tablehead{
\colhead{$\ell^\prime$}         &
\colhead{2}         &
\colhead{3}         &
\colhead{4}         &
\colhead{5}         &
\colhead{6}         &
\colhead{7}         &
\colhead{8}         &
\colhead{9}         &
\colhead{10}         &
\colhead{11}         &
\colhead{12}         &
\colhead{13}         &
\colhead{14}         &
\colhead{15}         &
\colhead{16}         &
\colhead{17}         &
\colhead{18}         &
\colhead{19}
}

\startdata
  0&   0&   0&   0&   0&   0&   0&   0&   0&   0&   0&   0&   0&   0&   0&   0&   0&   0&   0\nl
  1&   0&   0&   0&   0&   0&   0&   0&   0&   0&   0&   0&   0&   0&   0&   0&   0&   0&   0\nl
  2&\MS1087&   0&   0&   0&   0&   0&   0&   0&   0&   0&   0&   0&   0&   0&   0&   0&   0&   0\nl
  3&   0&\MS1016&   0& 275&   1&  84&   1&  21&   1&  15&   1&  16&   1&  12&   1&   7&   1&   5\nl
  4& 150&   0& 930&   0& 295&   1&  81&   1&  20&   1&  17&   1&  17&   1&  12&   1&   7&   1\nl
  5&   0& 197&   1&\MS1067&   0& 182&   1&  63&   1&  15&   1&   7&   1&   9&   1&   8&   1&   5\nl
  6&  35&   0& 251&   0&\MS1051&   1& 179&   1&  57&   1&  12&   1&   6&   1&   9&   1&   7&   1\nl
  7&   0&  46&   1& 138&   0&\MS1103&   1& 167&   1&  58&   1&  15&   1&   7&   1&   8&   1&   7\nl
  8&   6&   0&  55&   0& 143&   1&\MS1100&   1& 166&   1&  57&   1&  14&   1&   7&   1&   8&   1\nl
  9&   0&   9&   1&  38&   0& 132&   1&\MS1105&   1& 164&   1&  58&   1&  15&   1&   7&   1&   7\nl
 10&   3&   0&  11&   0&  37&   0& 134&   1&\MS1102&   1& 165&   1&  58&   1&  15&   1&   6&   1\nl
 11&   0&   5&   1&   7&   0&  38&   0& 135&   1&\MS1104&   1& 161&   1&  56&   1&  14&   1&   6\nl
 12&   3&   0&   8&   0&   6&   0&  39&   0& 139&   1&\MS1100&   1& 162&   1&  55&   1&  14&   1\nl
 13&   0&   5&   0&   3&   0&   8&   0&  41&   0& 138&   1&\MS1103&   1& 158&   1&  54&   1&  14\nl
 14&   2&   0&   7&   0&   3&   0&   9&   0&  42&   0& 140&   1&\MS1102&   1& 158&   1&  54&   1\nl
 15&   0&   3&   0&   3&   0&   4&   0&   9&   0&  41&   0& 138&   1&\MS1105&   1& 157&   1&  53\nl
 16&   1&   0&   4&   0&   4&   0&   4&   0&   9&   0&  42&   0& 139&   1&\MS1104&   1& 156&   1\nl
 17&   0&   2&   0&   3&   0&   4&   0&   4&   0&   9&   0&  42&   0& 139&   1&\MS1105&   1& 155\nl
 18&   1&   0&   2&   0&   3&   0&   4&   0&   4&   0&   9&   0&  42&   0& 139&   1&\MS1105&   1\nl
 19&   0&   1&   0&   1&   0&   3&   0&   4&   0&   4&   0&  10&   0&  42&   0& 139&   1&\MS1105\nl
\enddata
\end{deluxetable}

\begin{deluxetable}{ccccc}
\tablewidth{0pt}
\tablecaption{Binned power spectra of the 4 yr DMR Maps. \label{tab.logpower}}
\tablehead{
\colhead{$\ell$ Range}         &
\colhead{$\ell_{eff}$}         &
\colhead{NG $A \times B$}      &
\colhead{$53 \times 90$}       &
\colhead{$53+90\; A \times B$}
}

\startdata
2     &  \phn 2.1 & \phs$0.08 \pm 0.68$ & \phs$0.16 \pm 0.65 $ 
& \phs$0.15 \pm 0.60 $ \nl
3     &  \phn 3.1 & \phs$0.76 \pm 0.70$ & \phs$0.99 \pm 0.59 $ 
& \phs$0.90 \pm 0.57 $ \nl
4     &  \phn 4.1 & \phs$1.65 \pm 0.76$ & \phs$1.51 \pm 0.57 $ 
& \phs$1.52 \pm 0.56 $ \nl
5-6   &  \phn 4.6 & \phs$1.62 \pm 0.51$ & \phs$1.40 \pm 0.35 $ 
& \phs$1.30 \pm 0.34 $ \nl
7-9   &  \phn 6.3 & \phs$1.71 \pm 0.52$ & \phs$0.85 \pm 0.29 $ 
& \phs$0.94 \pm 0.28 $ \nl
10-13 &  \phn 8.9 &  $  -0.02 \pm 0.79$ & \phs$0.78 \pm 0.33 $ 
& \phs$1.19 \pm 0.30 $ \nl
14-19 &   11.5    & \phs$1.50 \pm 1.42$ & \phs$1.75 \pm 0.46 $ 
& \phs$1.52 \pm 0.40 $ \nl
20-30 &   13.2    & \phs$0.37 \pm 4.81$ & \phs$0.57 \pm 1.39 $ 
& \phs$0.02 \pm 1.08 $ \nl
\enddata
\tablecomments{All values have been normalized to the mean for 
$Q = 17\;\mu$K, $n = 1$ Monte Carlo runs.}
\end{deluxetable}

\begin{deluxetable}{ccc}
\tablewidth{0pt}
\tablecaption{Power Law Fits to 4 year DMR map power spectra.
\label{tab.Qn}}
\tablehead{
\colhead{Maps}         &
\colhead{$n_{app}$}    &
\colhead{$\Amp$ at $n_{app} = 1$, $[\mu$K]}
}

\startdata
53+90 A$\times$B &  \nTBmax    &  \Qmax    \nl
53$\times$90     &  \nTBcross  &  \Qcross  \nl
NG A$\times$B    &  \nTBng     &  \Qng     \nl
\enddata
\end{deluxetable}

\clearpage


\figcaption[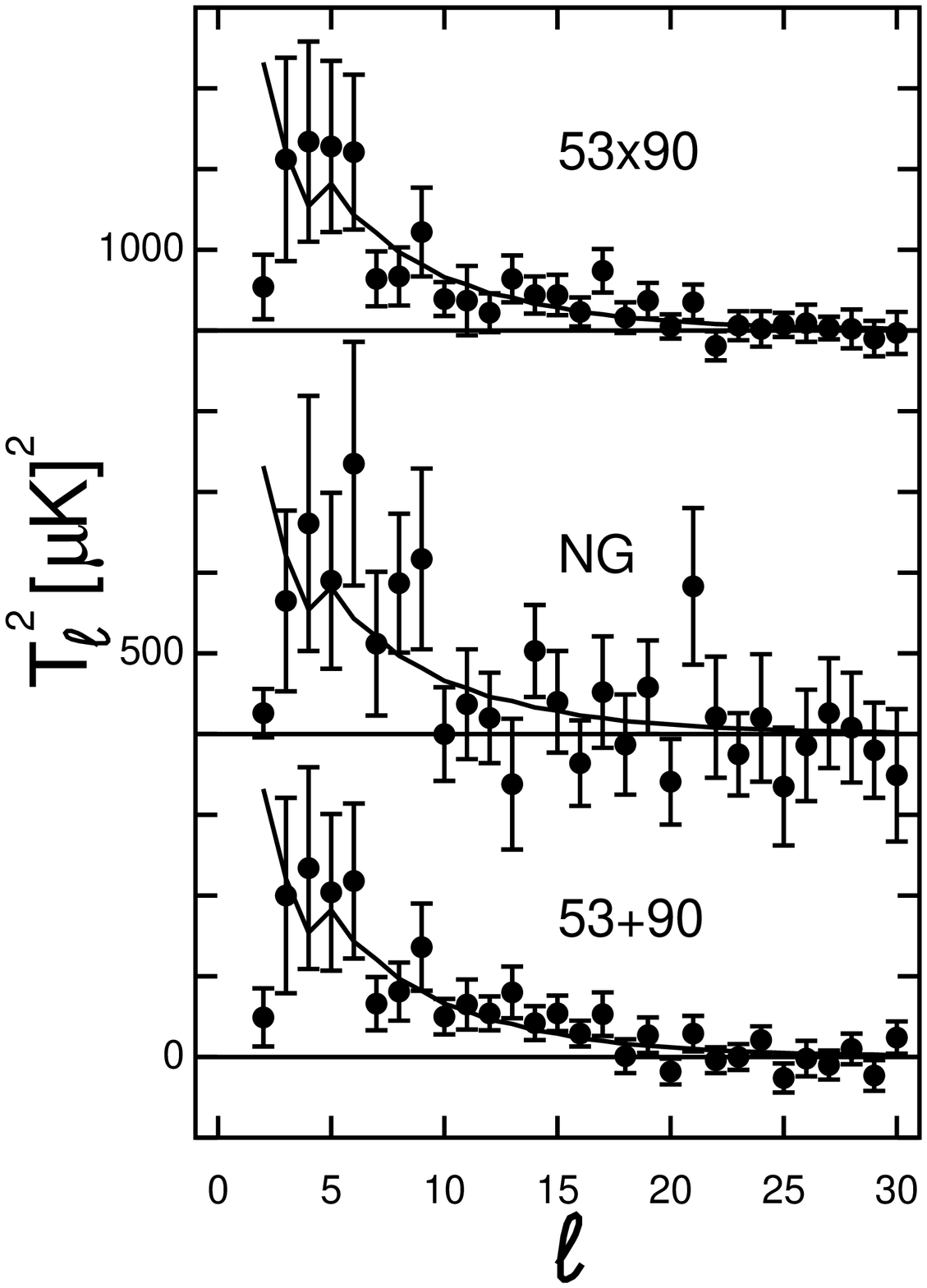]{Cross power spectra for the 53+90 $A\times B$,
$53 \times 90$ and NG $A\times B$ maps.
$T_\ell^2$ measures the variance of the sky due to order $\ell$ harmonics
for full sky coverage, but partial sky coverage changes the expected
value slightly as seen in the curves showing the average power spectrum
of $Q = 17\;\mu$K, $n = 1$ models in the cut sky.
Values are shifted upward by 400 for NG and 900 for $53 \times 90$,
as shown by the horizontal lines marking zero power.
\label{fig.linear}}

\figcaption[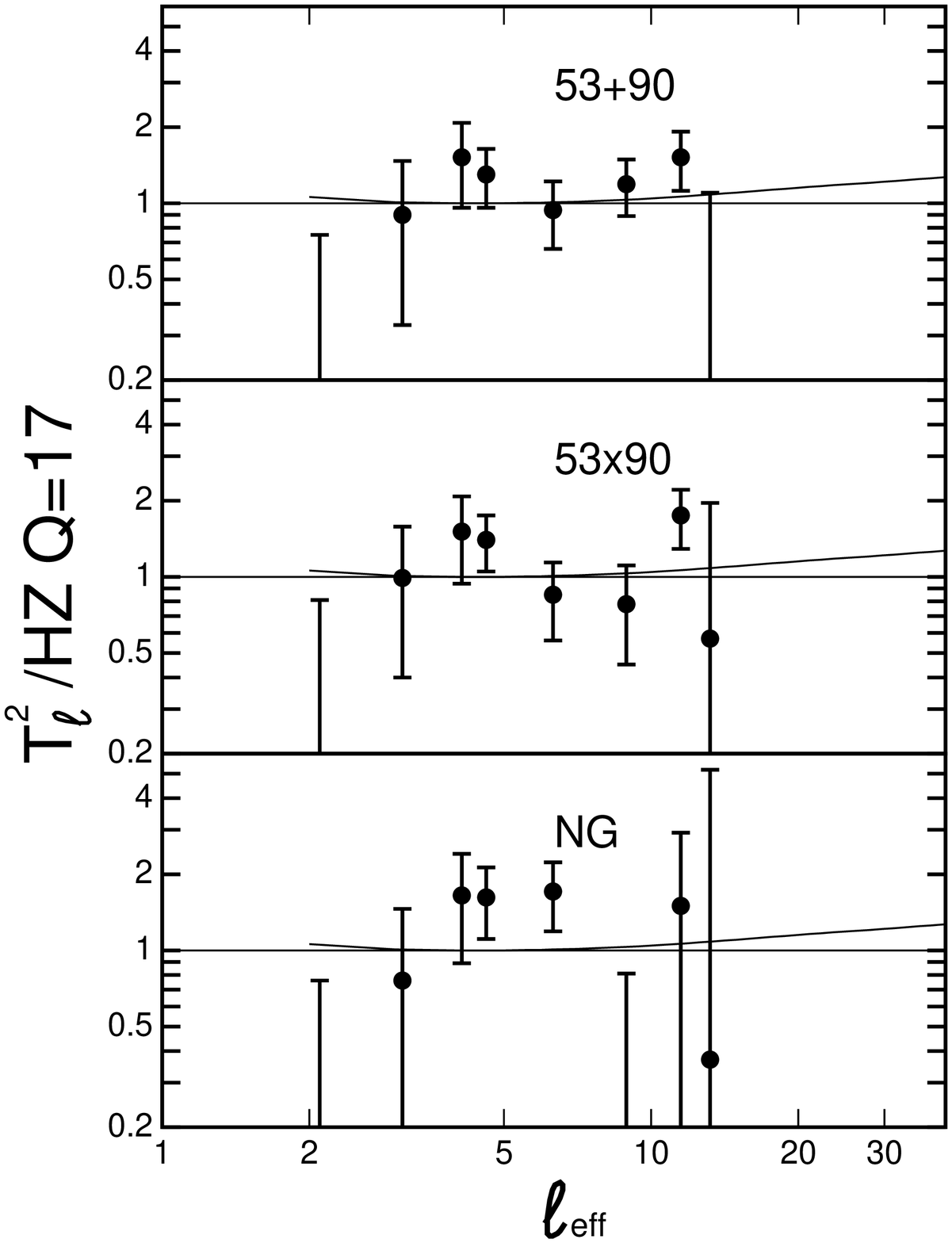]{Binned cross power spectra for the 53+90 $A\times B$,
$53 \times 90$ and NG $A\times B$ maps,
normalized to the mean power spectrum of $Q = 17\;\mu$K, $n=1$ simulations,
plotted on a logarithmic scale.
$\ell_{eff}$ is the effective wavenumber of the
bin for $n = 1$.
The thin curves show a CDM model with $n_{pri} = 0.96$ including the
effect of gravitational waves derived from 
Crittenden \etal\protect\markcite{CBDES93} (1993).
\label{fig.logpower}}

\figcaption[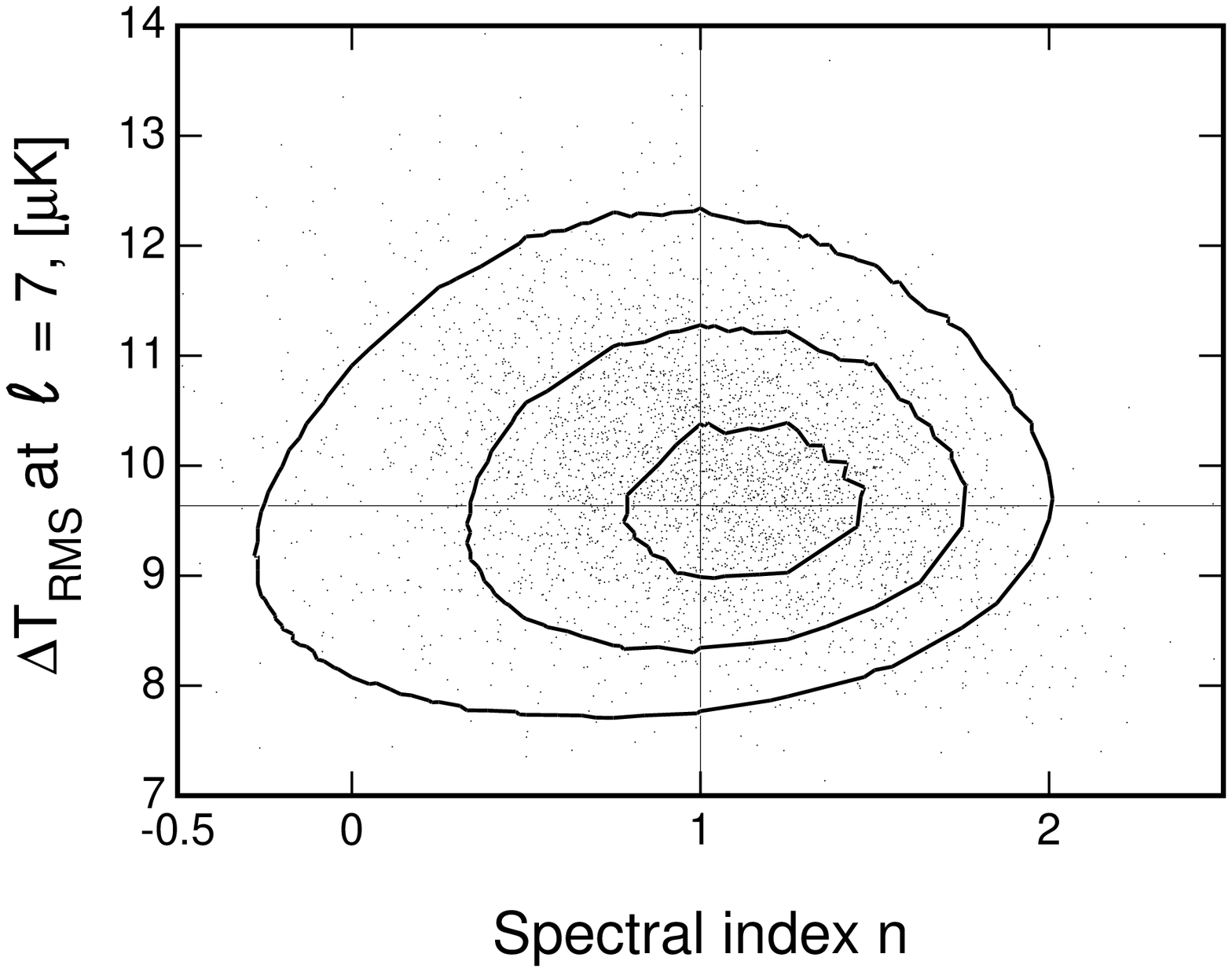]{Each point is an input parameter set that is 
consistent with the real 4 year $53 \times 90$ DMR data 
for a given realization of the random 
cosmic and radiometer variance processes.  
The amplitude is specified using the RMS $\Delta T$ due to $\ell = 7$
spherical harmonics because $\ell_{eff}$ for this fit is 7.3.
The likelihood contours 
are at $\Delta(-2\ln L) = 1$, 4 and 9.\label{fig.match}}

\setcounter{figure}{0}

\clearpage

\begin{figure}
\plotone{linear4y.ps}
\caption{~}
\end{figure}

\clearpage

\begin{figure}
\plotone{logpwr4y.ps}
\caption{~}
\end{figure}

\clearpage

\begin{figure}
\plotone{m5x9_4yr.ps}
\caption{~}
\end{figure}

\end{document}